\documentclass[11pt,twoside]{article}

\usepackage{asp2014}

\aspSuppressVolSlug
\resetcounters

\begin{document}

\title{The Real-Time Data Processor Framework for Data Handling and Analysis of High-Energy Instruments.}


\author{A. Bulgarelli,$^1$ N. Parmiggiani,$^1$ L. Castaldini,$^1$ R. Falco,$^1$ A. Di Piano,$^{1,2}$ V. Fioretti,$^1$ G. Panebianco,$^{1}$ A. Rizzo$^3$}
\affil{$^1$INAF/OAS Bologna, Via P. Gobetti 93/3, 40129 Bologna, Italy; \email{andrea.bulgarelli@gmail.it}}
\affil{$^2$Universit\`{a} degli Studi di Modena e Reggio Emilia, DIEF, Via Pietro Vivarelli 10, 41125 Modena, Italy;}
\affil{$^3$INAF/OA Catania, Via Santa Sofia 78, 95123 Catania, Italy.}

\paperauthor{Sample~Author1}{Author1Email@email.edu}{ORCID_Or_Blank}{Author1 Institution}{Author1 Department}{City}{State/Province}{Postal Code}{Country}
\paperauthor{Sample~Author2}{Author2Email@email.edu}{ORCID_Or_Blank}{Author2 Institution}{Author2 Department}{City}{State/Province}{Postal Code}{Country}
\paperauthor{Sample~Author3}{Author3Email@email.edu}{ORCID_Or_Blank}{Author3 Institution}{Author3 Department}{City}{State/Province}{Postal Code}{Country}



\begin{abstract}
We implemented a real-time data processor (rta-dp) framework that can be used to develop real-time analysis pipelines and data handling systems to manage high-throughput data streams with distributed applications in the context of ground and space astrophysical projects and high-energy instruments. The rta-dp is based on the ZeroMQ in-memory communication framework to receive input data, share data between distributed processes, and send or receive commands and pipeline configuration. The rta-dp framework has a flexible architecture that allows the implementation of distributed analysis systems customized to the requirements of several scenarios. The rta-dp framework also provides monitoring capabilities for the running processes and sends housekeeping, logging, alarms, and informative messages that a monitoring process can acquire. We are using the rta-dp in several contexts, such as acquiring and processing data from X-ray detectors to the data quality system of the ASTRI Project, as well as reprocessing and archiving data.
\end{abstract}



\section{Introduction}

Ground and space astrophysical projects and high-energy instruments facilities implement real-time data acquisition and analysis software to analyse the acquired data as soon as possible and generate scientific results. In this context, we implemented the real-time data processor (rta-dp) framework. This framework can be used to implement real-time analysis pipelines and data handling systems to manage high-throughput data streams with distributed applications.

\section{Software Architecture}

The real-time data processing system  is based on the ZeroMQ\footnote{https://zeromq.org/} in-memory communication framework to receive input data, share data between distributed processes, and send or receive commands and pipeline configuration. The rta-dp is available in both Python and C++ implementations. The pipelines implemented with the rta-dp are configured configured through a JSON\footnote{: https://www.json.org/} file.

With rta-dp, it is possible to build a data processing system as a chain of processes that analyses both a set of files in a batch mode or a data stream. 

Processes exchange data and messages categorized into data, commands, monitoring, alarms or events, configuration for workers, logs, and informative messages. Each message is formatted in JSON and includes a standard header containing metadata: type (to identify the message category), subtype (specific to the message type), time, source and target process of the message, priority, and body. The body holds the actual content of the message, which can vary based on its purpose. The body can contain the information, e.g. data. For data-related messages, subtypes define the format of the content. These include filename, where only the file name is transmitted without additional information; binary, which encodes arbitrary binary data; and string, allowing textual data in any format. The body field serves as a placeholder for any custom fields needed by each specific message type or subtype, ensuring flexibility and adaptability within the system.

\begin{figure*}[!htb]
	\centering
	  \includegraphics[width=1.0\textwidth]{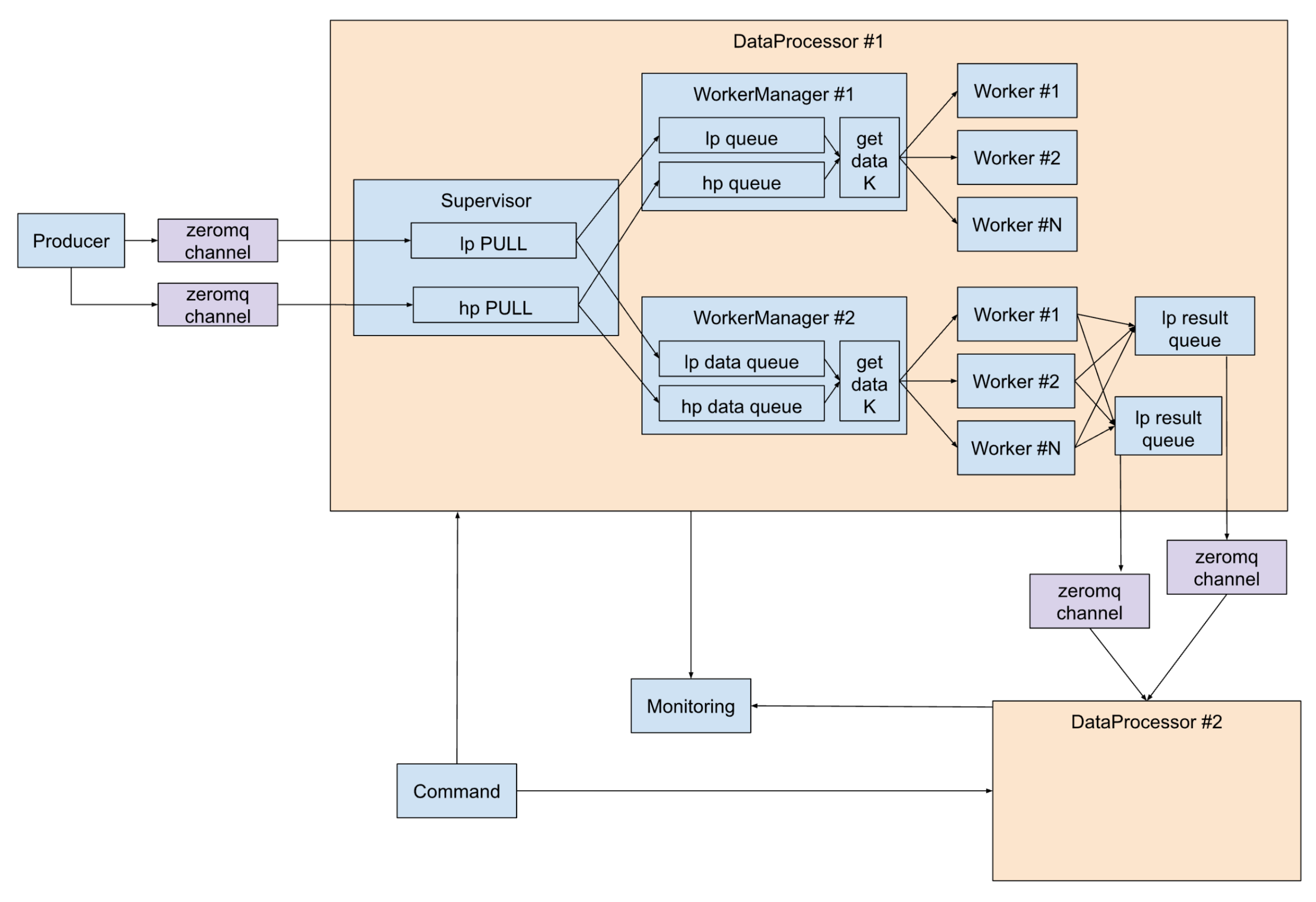}
	\caption{Software architecture. The figure shows a DataProcessor system where Producers send data via ZeroMQ PUSH/PUB channels to a Supervisor, which receives low-priority (lp) and high-priority (hp) tasks (files or single events in streaming). The Supervisor distributes tasks to WorkerManagers that manage queues and assign data to multiple Workers for processing. Workers send results back through ZeroMQ channels to another DataProcessor, completing the data flow. A Monitoring process collects monitoring information, and a Command process send data to DataProcessors}
	\label{fig:architecture}
\end{figure*}

A data processing chain is a set of Data Processors \ref{fig:architecture}. Each Data Processor can process the same input data in a parallel way and with different kinds of analysis, and a Supervisor manages it.

A Supervisor manages the Data Processor and receives the input data, managing low and high-priority data streams. The Supervisor receives data through messages via ZeroMQ, but it is possible to implement a custom data receiver, for instance, using Kafka\footnote{https://kafka.apache.org}, or to connect rta-dp Supervisor directly to a data acquisition system.

Each Supervisor instantiates one or more WorkerManagers following the configuration file and transfers the input data to them.

\begin{figure*}[!htb]
	\centering
	  \includegraphics[width=0.5\textwidth]{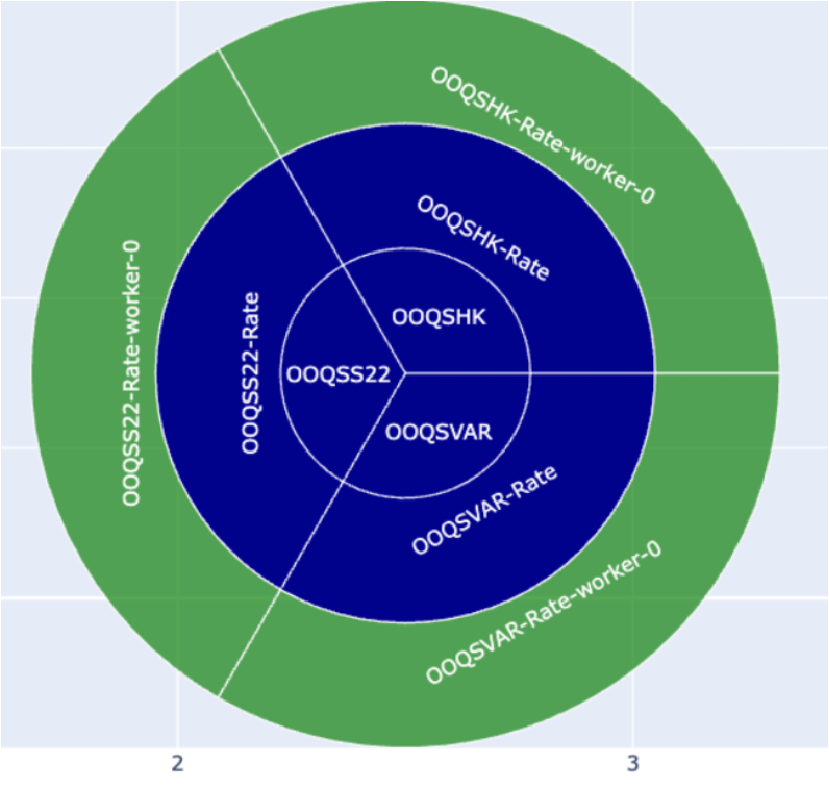}
	\caption{Monitoring GUI. The figure shows a monitoring layout for rta-dp with a hierarchical structure of metrics. At the center there are the worker state, surrounded by the state of the WorkerManagers, and finally the outermost layer includes the states of the Supervisors, indicating detailed monitoring at both global and worker levels.}
	\label{fig:mon}
\end{figure*}

Each Supervisor in the system manages communication through dedicated input and output channels. As input, a Supervisor connects to a command channel using the PUB/SUB pattern, where the Supervisor acts as the SUB client, and to a data channel that can use either the PUSH/PULL or PUB/SUB pattern. The type of data socket (push/pull or pub/sub) is specified in the configuration file.

As output, the Supervisor connects to channels for low-priority results (push/pull or pub/sub), and high-priority results (push/pull or pub/sub). 

Each Supervisor is permitted to collect monitoring points and forward them to an output monitoring queue for further processing. Monitoring points include, among others, the state of the processing (see Figure \ref{fig:mon}), the rates of data processing, and the number of packets in the queue. In addition, the rta-dp framework also provides additional monitoring capabilities for the running processes, sending logging, alarms, and informative messages that a monitoring process can acquire and analyse.

Each Supervisor can instantiate N WorkerManagers, one for each type of analysis that must be executed. The Supervisor collects and routes messages while WorkerManagers distribute tasks across Workers. All Workers handle the same tasks, but a datum is processed by only one Worker, ensuring load balancing. The WorkerManager instantiates one or more Workers using multithreading or multiprocessing features, as defined in the configuration file, to analyse the data in parallel. The Workers of the same WorkerManager execute the same analysis, getting the data from the low and high-priority queues and prioritising data queued in the high-priority flow. The analysis algorithms can be implemented externally to the rta-dp framework and included in the workers to separate the data analysis from the workflow management. 


A WorkerManager can send the analysis results as an output using ZeroMQ, which can become the input for a new rta-dp Supervisor. With this flexible architecture, it is possible to create a distributed analysis system customised to the requirements of each scenario. 

Each Supervisor, WorkerManager and Worker has a state machine. WorkerManager composes the state machines of the Workers, and Supervisor composes the state machines of the WorkerManagers.

\section{Conclusion}

We are using the rta-dp in several contexts, such as reprocessing, archiving data, and exploiting it in a real-time context, as well as the Online Observation Quality System \citep{parmiggiani_ooqs} of the ASTRI Project \citep{SCUDERI202252}. In addition, we are developing data handling systems for gamma-ray scintillator detectors using the rta-dp framework to exploit its flexibility and performance.


\bibliography{example}  

@INPROCEEDINGS{parmiggiani_ooqs,
       author = {{Parmiggiani}, N. and others},
        title = "{The New Architecture of the Online Observation Quality System for the ASTRI Mini-Array.}",
        booktitle = {Proceedings of the Astronomical Data Analysis Software and Systems XXXIV, 2024 (forthcoming)},
        year = {2024},
}

@article{SCUDERI202252,
title = {The ASTRI Mini-Array of Cherenkov telescopes at the Observatorio del Teide},
journal = {Journal of High Energy Astrophysics},
volume = {35},
pages = {52-68},
year = {2022},
issn = {2214-4048},
doi = {https://doi.org/10.1016/j.jheap.2022.05.001},
url = {https://www.sciencedirect.com/science/article/pii/S2214404822000180},
author = {S. Scuderi and others}
}


\end{document}